\documentclass[article,pra,a4paper]{revtex4}
\usepackage{graphicx}
\usepackage[normalem]{ulem}
\usepackage{natbib}
\bibpunct{(}{)}{;}{a}{,}{,}

\begin{document}
\author{ M. Wojciechowski and Marek Cieplak }
\affiliation{ Institute of Physics, Polish Academy of Sciences,
Al. Lotnik\'ow 32/46, 02-668 Warsaw, Poland}
\title{
Effects of confinement and crowding on folding of model proteins\\
{\small published in: Biosystems. 2008 Dec;94(3):248-52}
}

\begin{abstract}
We perform molecular dynamics simulations for a simple coarse-grained model of crambin placed inside of a softly repulsive sphere of radius $R$.
The confinement makes folding at the optimal temperature slower and affects the folding scenarios, but both effects are not dramatic.
The influence of crowding on folding are studied by placing several identical proteins within the sphere, denaturing them, and then by monitoring refolding. If the interactions between the proteins are dominated by the excluded volume effects, the net folding times are essentially like for a single protein. 
An introduction of inter-proteinic attractive contacts hinders folding when the strength of the attraction exceeds about a half of the value of the strength of the single protein contacts.
The bigger the strength of the attraction, the more likely is the occurrence of aggregation and misfolding.
\end{abstract}

\maketitle

\section{Introduction}
There is a growing interest in studies of biomolecules enclosed within a limited space. 
One reason is that almost all life processes take place in compartments such as cells where concentrations of proteins, lipids, shugars, and nucleic acids are large \citep{Ellis_2006}. 
Such conditions are also desired in artificial life systems such as the liposomes that allow for protein synthesis within their interior\citep{Murtas}.
Chaperonin cages \citep{Hartl_2002}, that assist in folding and refolding processes of proteins, offer an example of compartmentalization at a still smaller length scale. 
Another reason for the interest in the confinement effects is provided by recent advances in nanotechnology and resulting novel encapsulation techniques. 
These involve, for instance, reverse micelles which are mimetic systems of biological membranes composed of amphiphilic molecules. 
These molecules self-organize so that the polar head-groups point inward and hydrocarbon chains face the organic solvent \citep{Matzke,Luisi,Melo}.
The amount of the entrapped water is controlled by experimental conditions and a typical radius of the corresponding sphere can be as small as $\sim$20 {\AA}.
The water molecules at the inner surface have a propensity to organize \citep{Fayer} and the conditions within need not be uniform \citep{Levinger}.
When it comes to larger confined systems, there are many microfluidic ways to deposit droplets on surfaces, e.g. in the context of the protein and DNA microarrays \citep{Duroux}.

It is thus interesting to undertake theoretical studies of proteins that are confined. 
A simple way to introduce confinement is through a sphere \citep{Baumketner,Rathore_2006}, or a cage \citep{Takagi_2003}, which are repulsive to proteins located on the inside. 
A sphere which has  attractive hydrophobic and repulsive hydrophilic patches on the inside has been also discussed \citep{Jewett_2004} to elucidate the workings of chaperonins.
One can also generate cavities by using many spheres, repulsive on the outside, to immitate the effects of crowding \citep{Thirumalai_2005}.
Most of the studies carried out so far have been focused on thermodynamics.
The confinement has been found to lead to a greater thermodynamic stability, broader and taller specific heat and more compact unfolded conformations \citep{Takagi_2003,Rathore_2006}. 
Crowding is expected to enhance these effects even further \citep{Thirumalai_2005}.

\newpage
In this paper, we consider the kinetics of folding of a protein.
This problem has already been studied by, Baumketner et~al. \citep{Baumketner} and Jewett et~al. \citep{Jewett_2004}.
In the case of the confining repulsive sphere \citep{Baumketner}, the wall potential was represented by 
\begin{equation}
V_{wall,B}(r) = 4\varepsilon_{wall} \frac{\pi R_s}{5r}\left[ \left( \frac{\sigma}{r-R_s}
\right)^{10}-\left(\frac{\sigma}{r+R_s}\right)^{10} \right] \;,
\end{equation}
where $R_s$ is the radius of the sphere, $r$ is the distance of a C$^{\alpha}$ atom from the center of the sphere, $\varepsilon _{wall}$ is the strength of the potential, and $\sigma$ is take to be equal to $3.8\AA$, i.e. to the distance between two consecutive C$^{\alpha}$ atoms in a protein. 
The folding time, determined at various temperatures, has been found to depend on $R_s$ in a complicated manner.
For instance,  at temperatures below the optimal temperature it decreases with increasing the $R_s$, but it increases above this temperature.
In the case of the non-uniform sphere \citep{Jewett_2004} the physics involved depends on the strength of attraction to the hydrophobic patches in the model. 
If the attractive patches act as strongly  as the hydrophobic interactions in the protein, the protein sticks to the wall and folding is arrested, i.e. it takes forever. 
A reduction in the strength of the attraction leads to a lowering of the folding time until a minimum is reached and then the folding time increases to a finite value when the wall becomes purely repulsive. 

Here, we focus on the effects of confinement and crowding within a softly repulsive sphere. 
We investigate the kinetics of folding of a single protein as a function of the radius of the sphere and then also as a function of the interactions between atoms belonging to different proteins when several proteins are confined together.
In order to house the proteins, we consider a wall represented  by the truncated and shifted Lennard-Jones potential
\begin{equation}
V_{wall}(r)  = \left\{ \begin{array}{ll}
4\varepsilon_{wall} \left[ \left( \frac{\sigma}{R-r}\right)^{12}-
\left(\frac{\sigma}{R-r}\right)^6 \right] +\varepsilon_{wall} & \;\;\;\;\;for \;\;(R-r) < r_0 \\
0 & \;\;\;\;\; for \;\; (R-r)\geq r_0 
\end{array} \right.
\end{equation}
where $R$ will be referred to as the radius of the sphere, and $\sigma = r_0\cdot 2^{-\frac{1}{6}}$.
We take $r_0$= 4 {\AA} (which is equal to the size of the repulsive core in the non-native contacts as defined below).
The specific form of a purely repulsive potential representing the sphere should not matter much for the kinetics of folding. 
However, what may matter more is the choice of a model of the protein.
Baumketner et al. \citep{Baumketner} use a 27-bead minimalistic model proposed by Honeycutt and Thirumalai \citep{Honeycutt} in which pair-wise interactions depend on whether the amino acids involved are hydrophobic or polar.

We use another simple coarse-grained model, a Go-like model \citep{GoAbe1_1981}, in an implementation developed in refs. \citep{GO_Cieplak_2000,GO_Cieplak_2003,GO_Cieplak_2004,GO_Cieplak_2005,Sulkowska_2005,Szymczak_2006} and perform molecular dynamics studies of folding and unfolding.
The Go-like models are rather imperfect tools to use in the context of folding, but are often found to be adequate to settle various qualitative issues, especially of a comparatory nature.
Their advantage is that they allow for a thorough statistical analysis of time dependent processes involving large conformational changes.

\begin{figure}[ht]
\par\centering
\resizebox*{0.9\textwidth}{!}
{\includegraphics[width=0.90\textwidth]{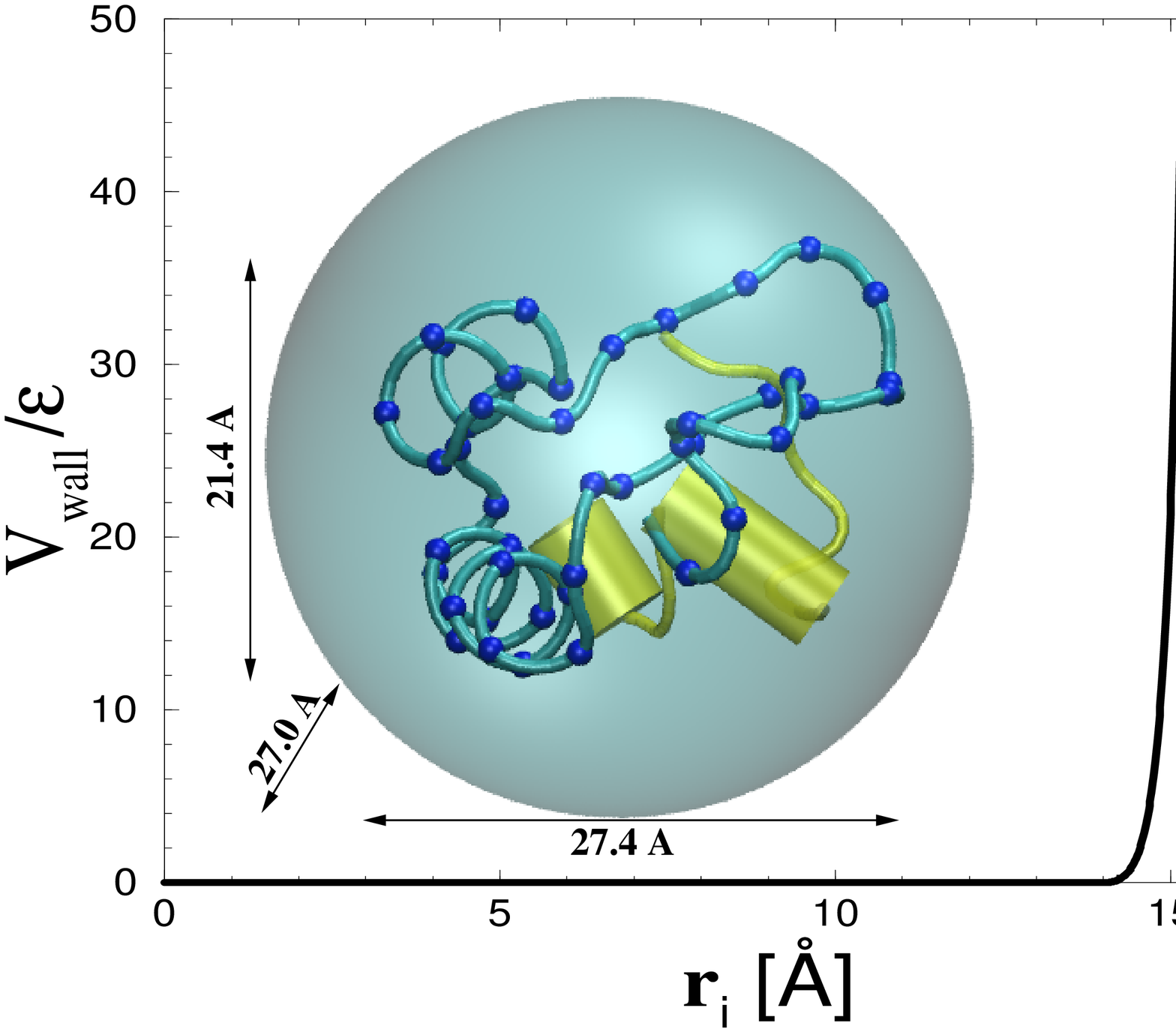}}
\par
\caption{
The potential of the spherical wall with $R$=18 {\AA} used in this paper.
$r$ denotes the distance of a bead form the center of the sphere.
}
\label{fig1}
\end{figure}

As an illustration of this approach, we consider crambin.
This is an $\alpha-\beta$ protein comprising of 46 amino acids. 
In its native state, the radius of gyration is about 9.7 {\AA} and the largest distance between a pair of its C$^{\alpha}$ atoms is 30 {\AA}.
The minimum value of $R$ that does not violate the steric constraints and still allows for meaningful conformational transition is 18 {\AA}
and the corresponding plot of the $V_{wall}$ potential is shown in Figure \ref{fig1}, together with a schematic representation of the native conformation.

We find that confinement under optimal folding conditions makes folding last longer but not more than by a factor of 2. 
We then consider the effects of crowding by placing up to twelve identical proteins inside of the sphere and studying refolding of the thermally denatured conformations. 
If a protein acts on another protein only through their excluded volumes, then the folding process, at optimality, is almost the same as for the single protein case. 
If one introduces attractive contacts between the proteins then, above their certain strength, the folding is hindered more substantially, even under the optimal conditions, since the conformational collapse now competes with aggregation.

\section{Methods}

The details of the approach are explained in refs. \citep{Sulkowska2007,Sulkowska2008}.
Briefly, each amino acid is represented by a bead located at the C$^{\alpha}$ position. 
The beads are tethered into a chain by the harmonic interactions. 
The local backbone stiffness is represented by a chirality potential \citep{GO_Cieplak_2005} that favors the native sense of the chirality, i.e. the native values of the dihedral angles.
The interactions between the amino acids are divided into the repulsive non-native contacts and attractive native contacts. 
The division is based on the absence or presence of the atomic overlaps in the experimentally determined native conformation. 
The attractive native contacts between amino acids $i$ and $j$ in distance $r_{ij}$ are described by the Lennard Jones potential $V^{LJ}_{ij}=4\varepsilon ((\frac{\sigma_{ij}}{r})^{12}-(\frac{\sigma_{ij}}{r})^6)$,
where the length parameter $\sigma_{ij}$ is determined so that the minimum in the potential agrees with the distance determined experimentally. 
The energy parameter $\varepsilon$ should be of order 1 - 1.6 kcal/mol.
We set $\varepsilon _{wall}$ to be equal to $\varepsilon$.
The model incorporates implicit solvent effects through temperature controlled random forces and strong velocity dependent damping.
Room temperature situations are considered to arise in the vicinity of $k_BT/\varepsilon$ of order of 0.3 ($k_B$ is the Boltzmann constant and $T$ is the temperature).
The time scale, $\tau$, in the molecular simulations is of order of 1 ns as it is set by time needed to cover distances of order of a typical $\sigma _{ij}$ through diffusion and not through a ballistic motion.

When studying folding, we determine the median time, $t_{fold}$, needed to establish all native contacts for the first time when starting from an unfolded conformation. 
A native contact is assumed to be established when 
$ r_{ij} < 1.5 \; \sigma_{ij}$.
$t_{fold}$ is determined based on at least 301 trajectories.
A convenient way to represent the sequencing of the folding events is through the scenario diagrams in which one shows average times when a specific contact gets established for the first time. 
The contacts are labelled by their sequential distance $|i-j|$.

Simulations of folding require a prior generation of extended denatured structures to fold from. 
These are obtained by unfolding the native structure by applying a high temperature.
Time scales required to arrive at a state in which all native contacts are broken are too long to achieve in the simulations. 
Instead, we take the criterion of all contacts with $|i-j| > 5$ being broken (see a discussion in ref. \citep{Sulkowska2007}). 
The corresponding median unfolding time is denoted by $t_{unf}$.

Confinement puts restrictions on the possible starting conformations since they must fit the sphere.
Thus we generate the starting sets by first placing a protein (or proteins) in a native state in a sphere of radius $R_0$, then setting $k_BT/\varepsilon$ at 1.0, and finally by storing conformations obtained at the end of a 10000$\tau$-long process of the unfolding dynamics (the corresponding $t_{unf}$ ranged between 100 and 2000 $\tau$ depending on $R_0$).
One can simulate refolding either for $R=R_0$, which is the simplest situation, but one may also consider refolding of the $R_0$-generated structures in a larger space when $R> R_0$ and, in particular, for $R=\infty$.

\begin{figure}[ht]
\par\centering
\resizebox*{0.9\textwidth}{!}
{\includegraphics[width=0.5\textwidth]{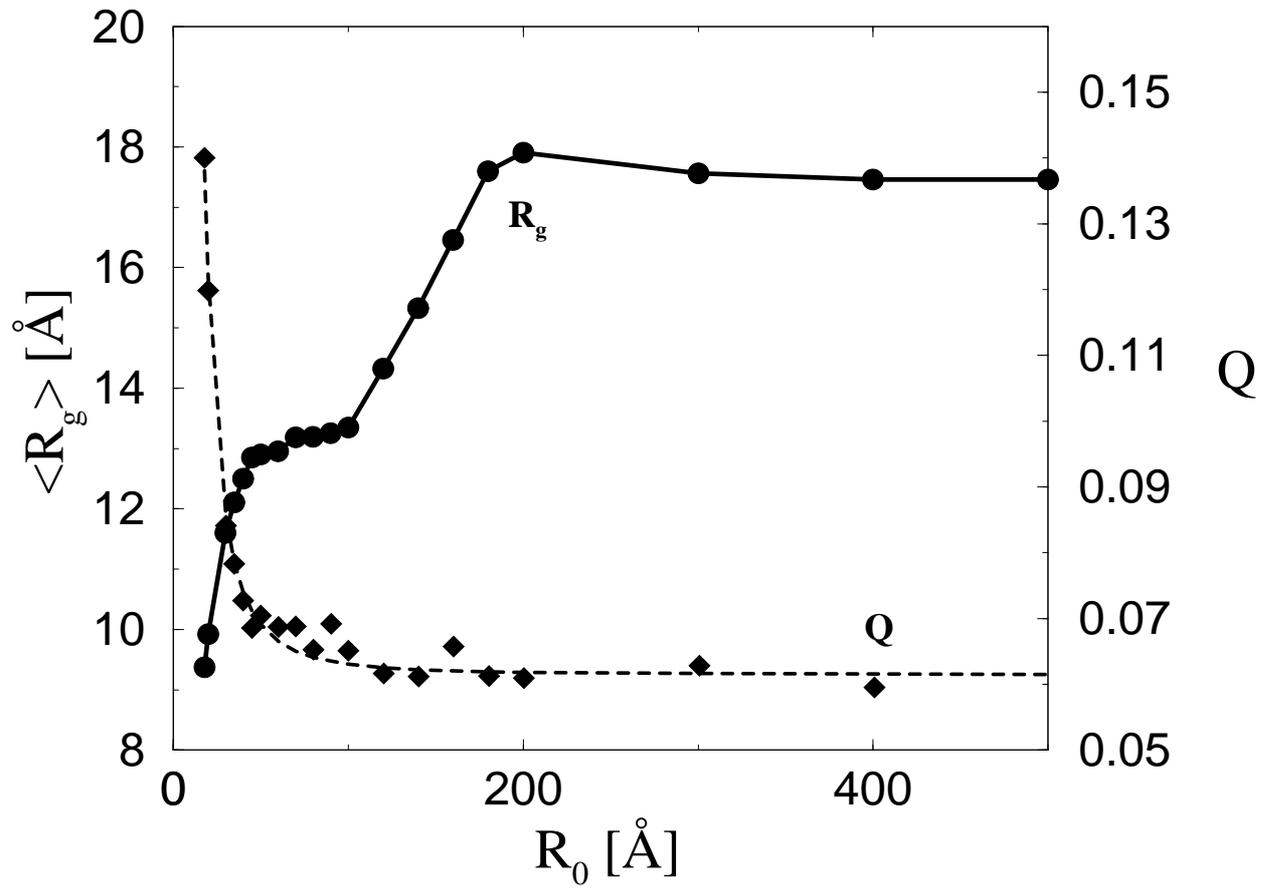}}
\par
\caption{
Geometrical characteristics of the unfolded structures of a single protein, as obtained for $k_BT/\varepsilon$=1.0, as a function of the radius of the sphere in which they were generated.
}
\label{fig2}
\end{figure}

The resulting unfolded structures are governed by the value of $R_0$ as shown in Figure~\ref{fig2} which provides structure characterization through the average radius of gyration, $<R_g>$, and the average fraction of the native contacts that are still present, $Q$.
Notice that even for very large values of $R_0$ there is always a small fraction of unbroken native contacts. 
These are usually associated with the $\alpha$ helices.

\begin{figure}[ht]
\par\centering
\resizebox*{0.9\textwidth}{!}
{\includegraphics[width=0.4\textwidth]{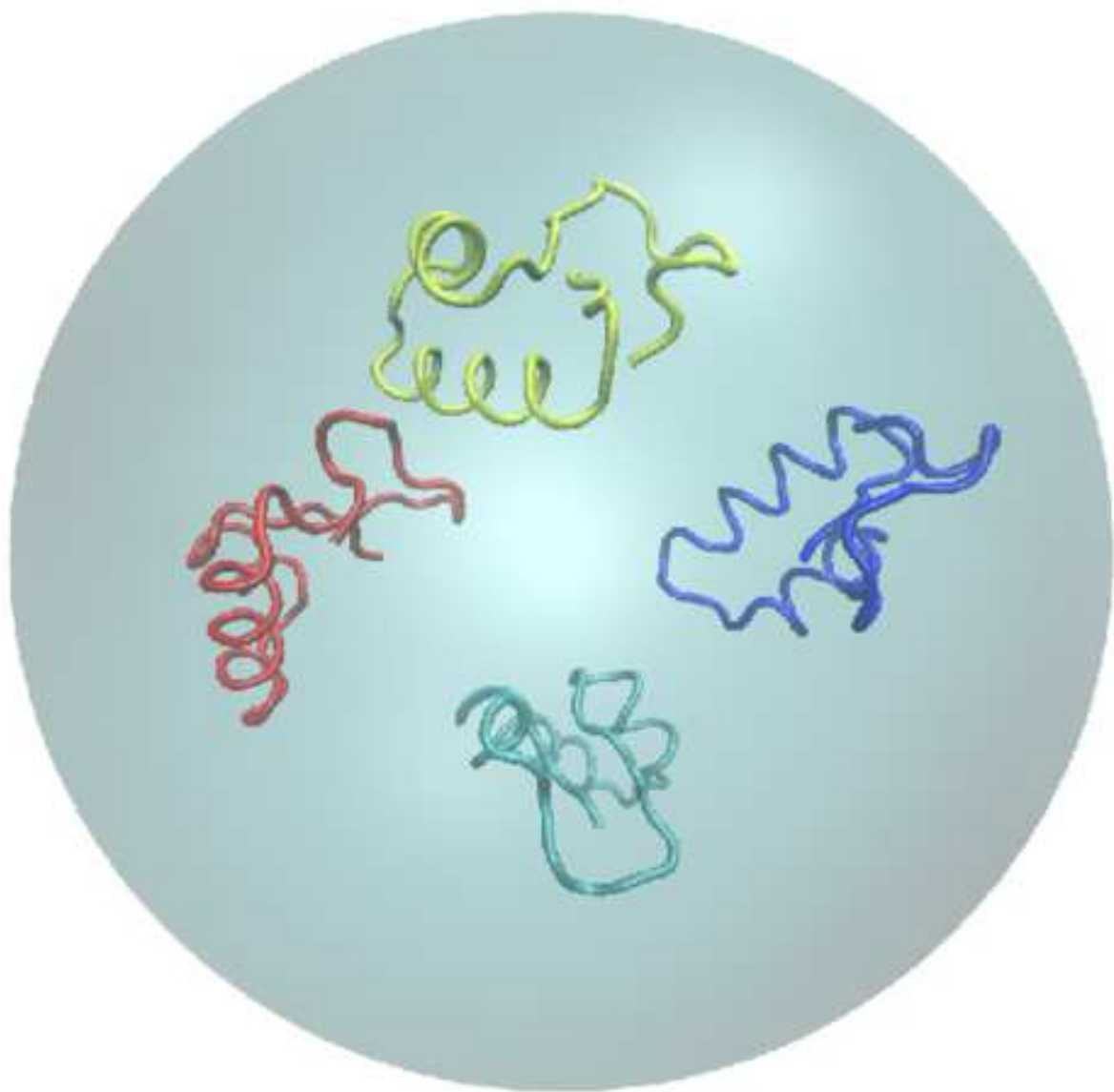}}
\par
\caption{
Four model molecules of crambin in their native state placed in the sphere.
}
\label{fig3}
\end{figure}

\begin{figure}[ht]
\par\centering
\resizebox*{0.9\textwidth}{!}
{\includegraphics[width=0.4\textwidth]{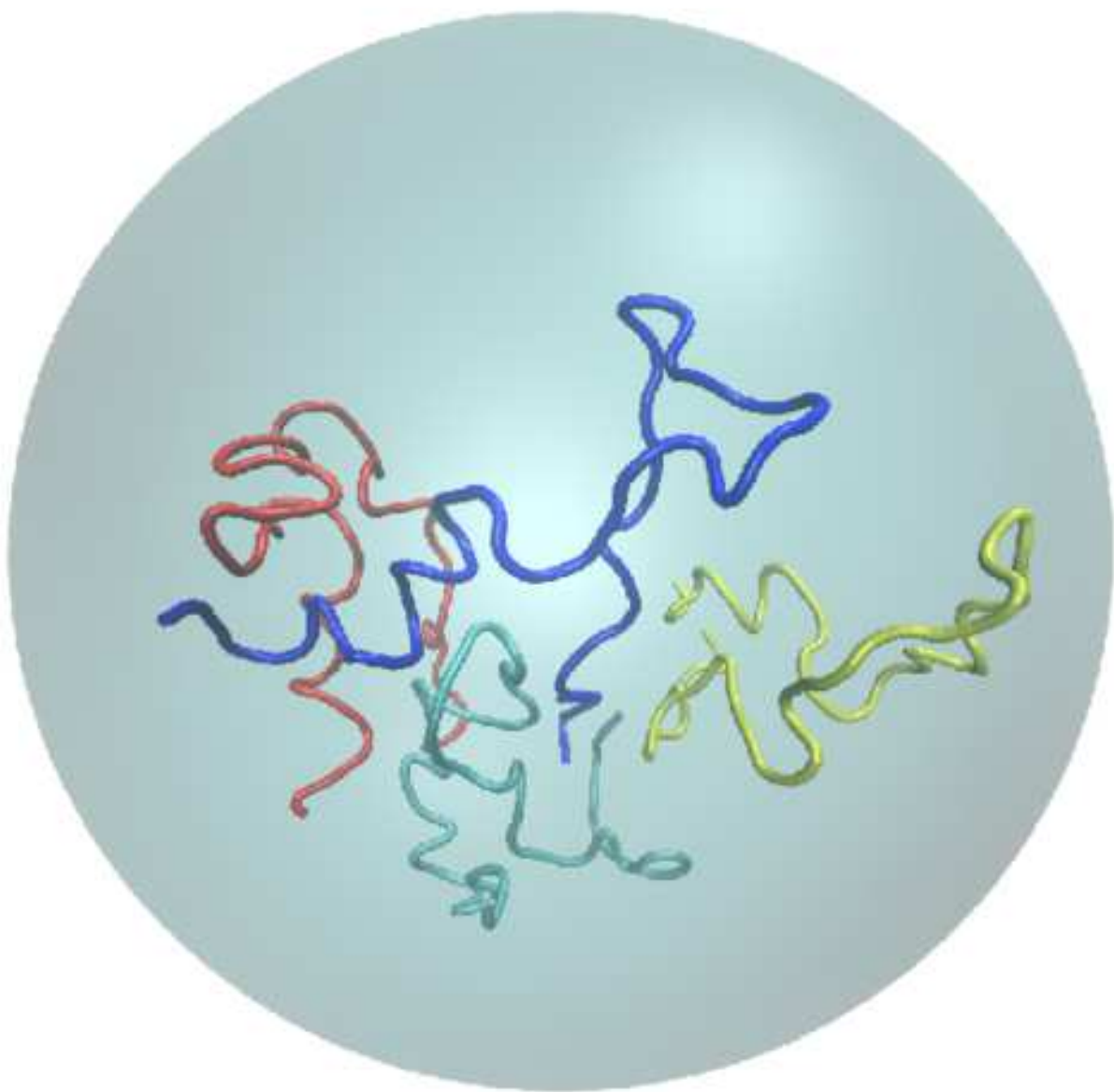}}
\par
\caption{
An example of an unfolded conformation of four proteins.
}
\label{fig4}
\end{figure}

In the case of $n>1$ molecules ($n$ up to 12 was considered), we place them together at the center of the sphere and then move them away from one another in small steps along arbitrary directions until they stop overlapping (see Figure \ref{fig3}). 
The system is then unfolded thermally for 10000 $\tau$ and the resulting structures (see Figure \ref{fig4}) are used for refolding studies.
$t_{fold}$ for $n$ molecules was determined by calculating the folding time of the individual molecules and then by taking the median over the molecules and over various trajectories. 
Each trajectory was stopped either when a time cutoff was exceeded or when each molecule was declared to arrive at the native conformation at some point during the evolution.
Another possible criterion would involve requiring a simultaneous establishment of all native contacts in the system. 
This would yield folding times that are significantly longer 
%$n$ times longer 
and comparing systems with different values of $n$ would not be relevant in the context of crowding since the behavior of a single molecule is what is of the interest here.
The scenarios of folding were realized like in the case of one molecule -- just the averaging gets enhanced by the factor of $n$.

\section{Folding of a single protein}

\begin{figure}[nt]
\par\centering
\resizebox*{0.9\textwidth}{!}
{\includegraphics[width=0.95\textwidth]{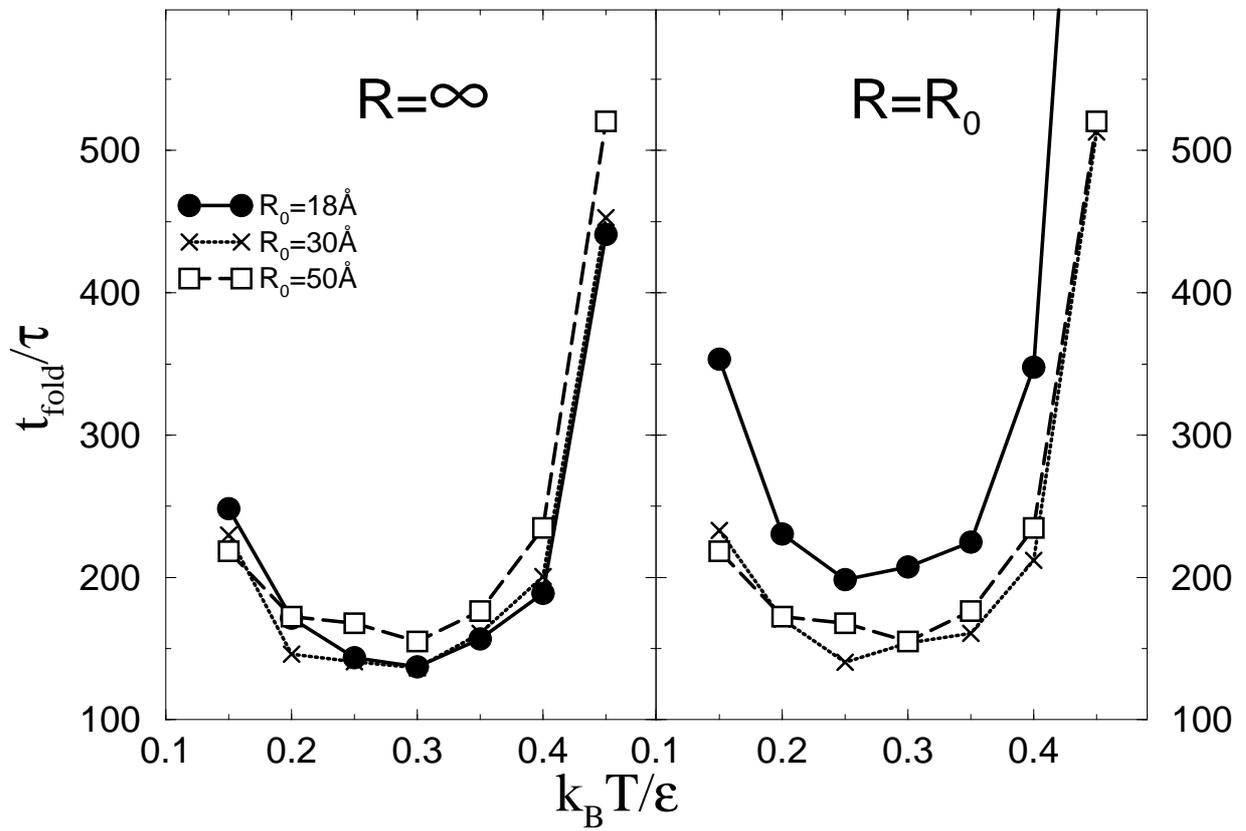}}
\par
\caption{
The median folding times for the model crambin as described in the main text.
The left panel is for folding in an inifinitely large sphere. 
The values of $R_0$ characterize the nature of the starting unfolded conformations.
The right panel is for spheres of three sizes as indicated.
}
\label{fig5}
\end{figure}

Figure \ref{fig5} shows $t_{fold}$ as a function of $T$ for the three choices of $R_0$: 18, 30, and 50 {\AA}.
The left panel corresponds to refolding in an unrestricted space whereas the right panel to refolding in the sphere with the original radius of confinement.
In the unrestricted space, $t_{fold}$ is nearly independent of $R_0$.
On the other hand, the presence of the sphere makes a change.
There is not much of a difference in folding between $R_0=R=30$ {\AA} and $50$ {\AA}, but there is a noticeable increase in $t_{fold}$ as one considers smaller values of $R$: $t_{fold}$ increases from 150 to 200 $\tau$ when $R$=18 {\AA} is considered.
The differences become more visible when comparing folding from starting conformations corresponding to a given $R_0$ and then evolved in the sphere or in the unrestricted space. For $R_0$=18 {\AA}, the change is from 137 $\tau$ to 200 $\tau$, i.e. by a factor of 1.5.
We interpret this finding as originating from the wall exerting a restriction on the process of the collapse: the protein may bounce with the wall on the way to its globular form and may need time to find another path.

\begin{figure}[ht]
\par\centering
\resizebox*{0.9\textwidth}{!}
{\includegraphics[width=0.95\textwidth]{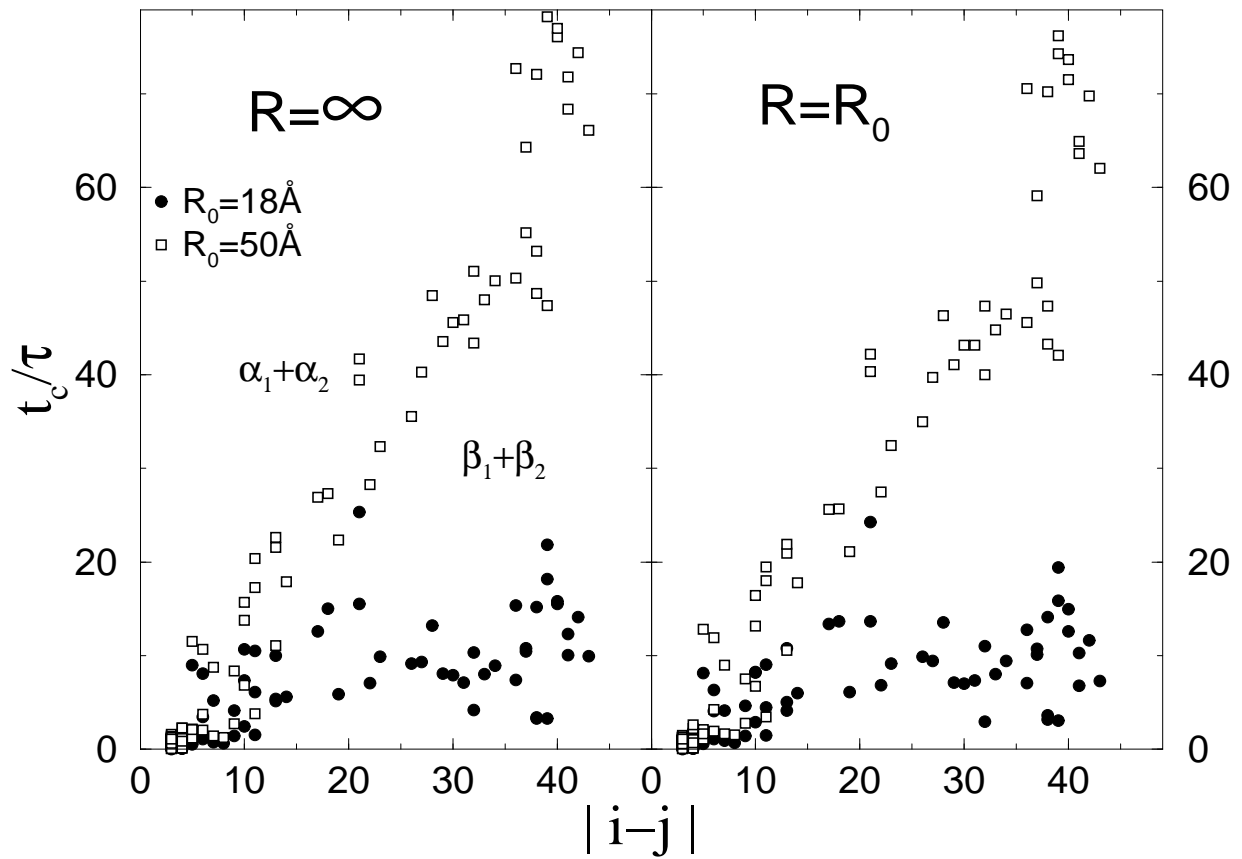}}
\par
\caption{
The folding scenarios in the unrestricted space (the left panel) and confined space (the right panel) for $R_0$=18 {\AA} (the solid symbols) and 50 {\AA} (the open symbols).
The temperature is set to $k_BT/\varepsilon=0.3$ which corresponds to the temperature of optimal folding.
The symbols $\alpha_1+\alpha_2$ and $\beta_1 + \beta_2$ indicate contacts between the two $\alpha$-helices and two $\beta$-strands that are present in crambin.
}
\label{fig6}
\end{figure}

Figure \ref{fig6} shows that the folding scenarios show more substantial sensitivity to the value of $R_0$ of the starting conformations than the folding times do (both for the unrestricted refolding, shown in the left panel, and for $R=R_0$).
The observation is that the tighter the initial confinement, the faster the establishment of the individual native contacts.
The complete folding, however, requires a simultaneous establishment of all contacts and this circumstance is sensitive to confinement to a much lesser degree.

\section{Folding of a several proteins}

\begin{figure}[ht]
\par\centering
\resizebox*{0.9\textwidth}{!}
{\includegraphics[width=0.95\textwidth]{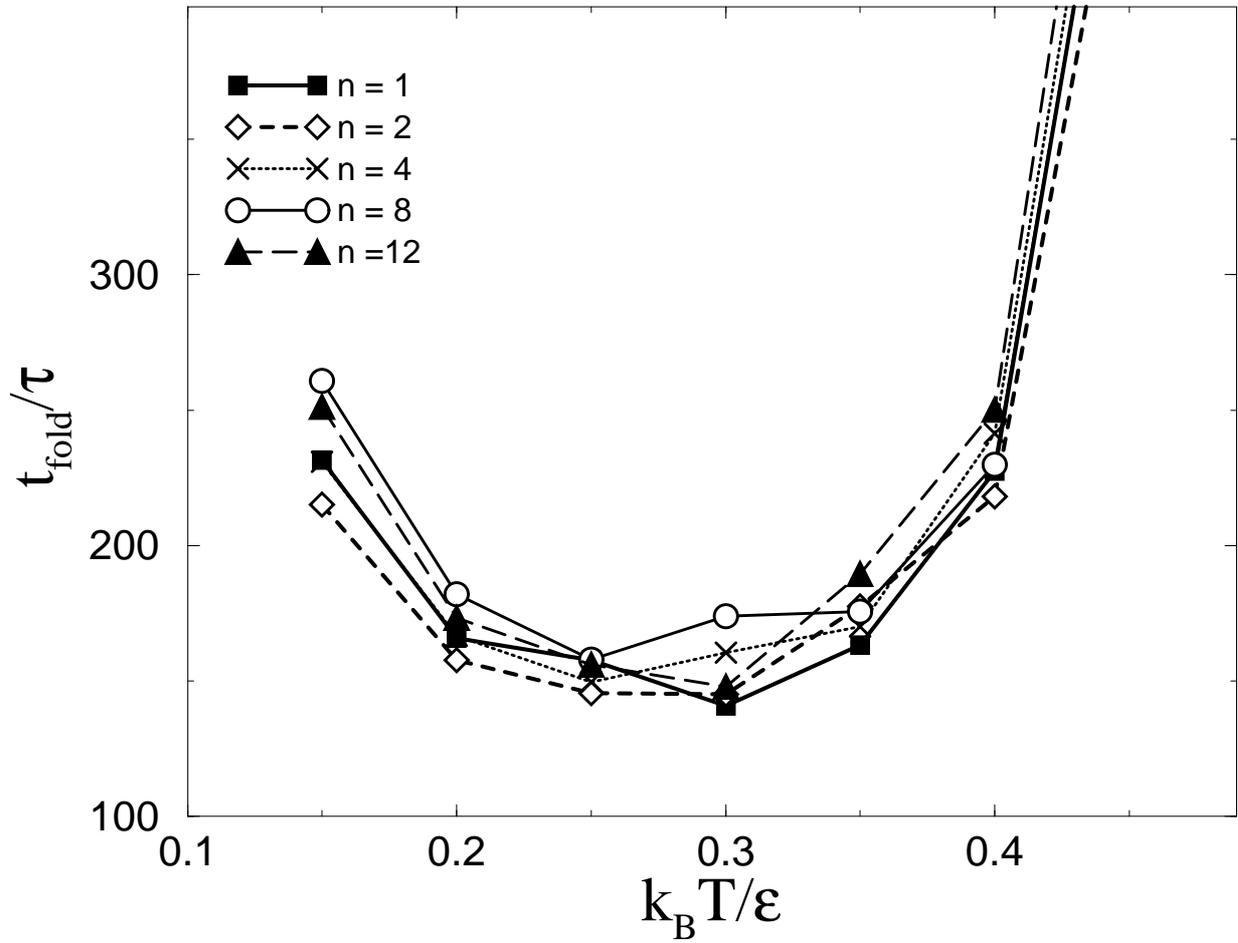}}
\par
\caption{ Folding time for an ensemble of $n$ proteins in the spherical cavity of $R$=36 {\AA} when there are no attractive interactions between the proteins. 
The values of $n$ range from 1 to 12 as indicated.
}
\label{fig7}
\end{figure}

We now discuss the effects of crowding and we place $n$ crambin molecules together into a sphere of radius \mbox{$R$=36 {\AA}.}
$n$ varies between 1 and 12.
We monitor the folding process when $R$=$R_0$. 
We first consider the simplest case in which the only way one molecule knows about the other is through the finite size of the hard cores ($r_0=4$ {\AA}) associated with the beads.
Fig~\ref{fig7} shows that under the conditions of optimal folding, $t_{fold}$ essentially does not depend on $n$, i.e. when folding goes well, processes in one molecule do not affect movements of the other.

We now repeat these calculations after introducing attractive interactions between the molecules. 
Since these molecules are all identical, the simplest way to do it is by introducing inter-protein attractive contact interactions in the following way. 
If there is a contact between $i$ and $j$ in one molecule $\lambda$ then we generate a similar contact between $i_{\lambda}$ in molecule $\lambda$ and $j_{\kappa}$ in molecule $\kappa$.
Each intermolecular contact is assigned an amplitude of $\varepsilon _I$ in the corresponding Lennard-Jones potential.
We vary $\varepsilon _I/\varepsilon$ between 0 and 1. 
The value of $\sigma _{ij}$ is kept the same as for the single protein.
The corresponding potential is given by
\begin{equation}
V_I(r_{ij'})  = 
\left\{ \begin{array}{lll}
	4\varepsilon \left[ \left( \frac{\sigma_{ij}}{r_{ij'}}\right)^{12}-
							\left(\frac{\sigma_{ij}}{r_{ij'}}\right)^6 
					\right] +	\varepsilon	-	\varepsilon_I &
\;\;\;\;\;for \;\;(r_{ij'}) < 2^{1/6} \; \sigma_{ij} & \;\; (repulsive) \\
	4\varepsilon_I \left[ 	\left( \frac{\sigma_{ij}}{r_{ij'}}\right)^{12}-
								\left(\frac{\sigma_{ij}}{r_{ij'}}\right)^6 \right] &
\;\;\;\;\; for \;\; r_{ij'}\geq 2^{1/6} \; \sigma_{ij} & \;\; (attractive)
\end{array} \right.
\end{equation}
where the prime in $j'$ indicates a different molecule.

\begin{figure}[ht]
\par\centering
\resizebox*{0.9\textwidth}{!}
{\includegraphics[width=0.95\textwidth]{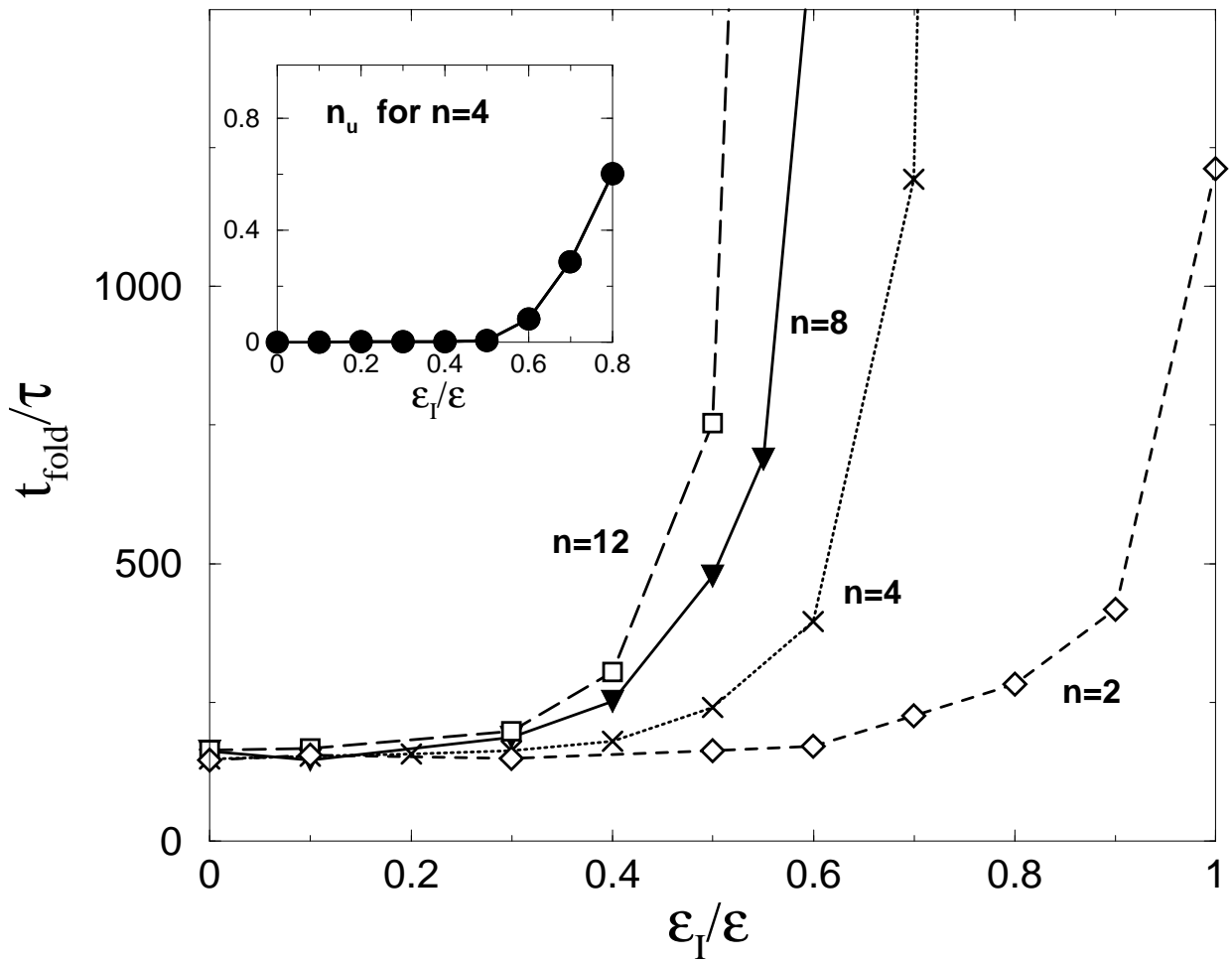}}
\par
\caption{
Median folding time for $n$ proteins at $k_BT/\varepsilon$=0.3 as a function of the inter-protein interaction strength $\varepsilon _I$.
The inset shows the fraction of misfolded trajectories.
}
\label{fig8}
\end{figure}

\begin{figure}[ht]
\par\centering
\resizebox*{0.9\textwidth}{!}
{\includegraphics[width=0.95\textwidth]{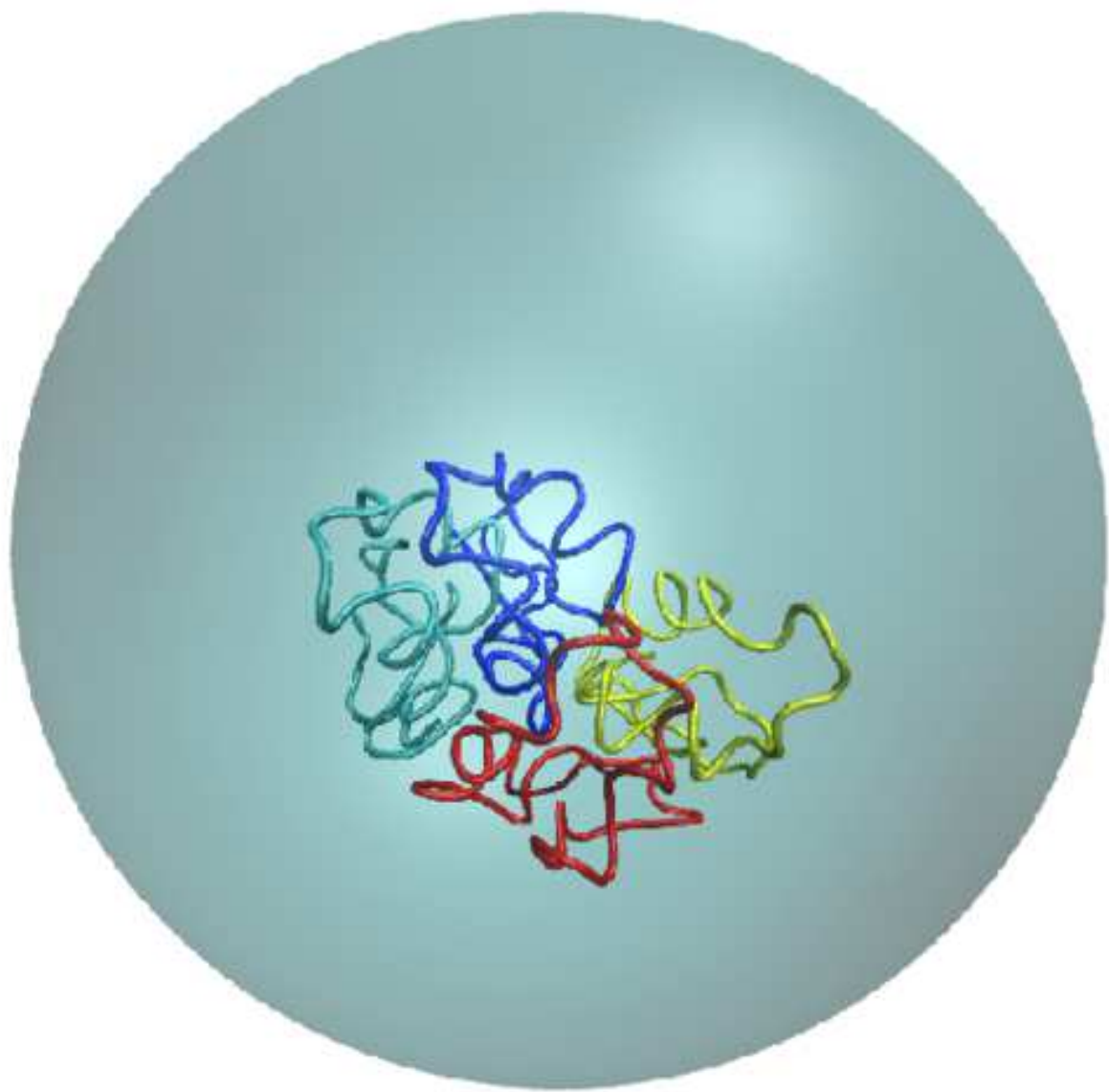}}
\par
\caption{
An example of a properly folded ensemble of four proteins for $\varepsilon_I/\varepsilon = 0.5$.
}
\label{fig9}
\end{figure}

The resulting folding times are shown in Figure \ref{fig8}.
It is seen that $t_{fold}$ is not affected by the inter-protein attraction as long as $\varepsilon _I/\varepsilon$ does not exceed a treshold value and then essentially all trajectories fold.
An example of a folded conformation is shown in Figure \ref{fig9}. 
For larger values of $\varepsilon _I$, the molecules influence one other which makes $t_{fold}$ longer and increases the frequency of the misfolding events, as shown in the inset of Figure \ref{fig8}.
The threshold value decreases with $n$. For $n$=4 it is $\sim 0.5$ and for $n$=12 it is $\sim 0.3$. 

\begin{figure}[ht]
\par\centering
\resizebox*{0.9\textwidth}{!}
{\includegraphics[width=0.95\textwidth]{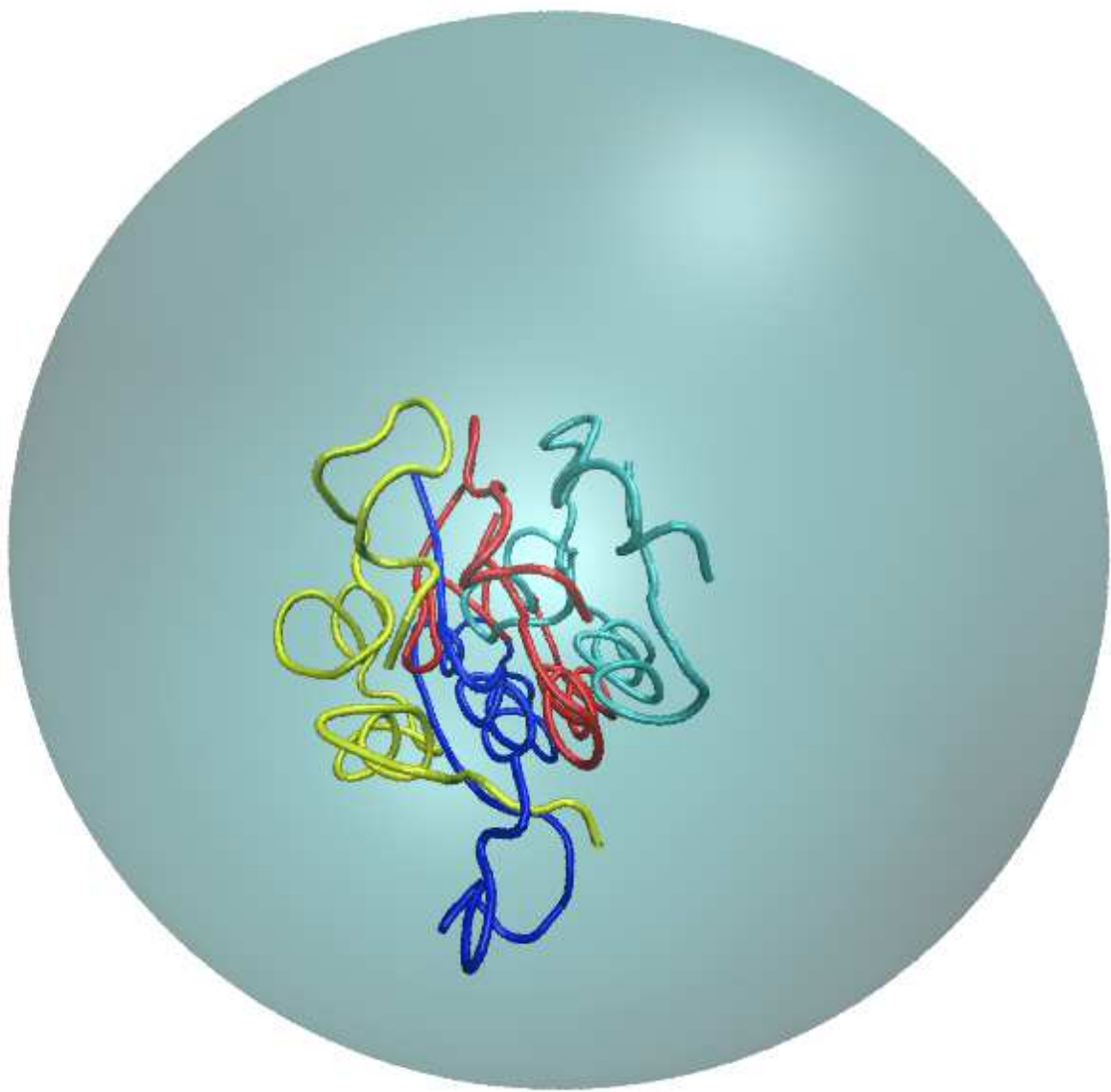}}
\par
\caption{
An example of a misfolded and entangled conformation of four proteins corresponding to $\varepsilon_I/\varepsilon = 0.8$.
}
\label{fig10}
\end{figure}

The misfolding is primarily due to aggregation and entanglement.
An example of a misfolded conformation is shown in Figure \ref{fig10}.
In most trajectories, the secondary structures (like helices 7-19 and 23-30) in individual molecules are established the first. 
At this stage, misfolding may occur since the intermolecular interactions tend to bring the secondary structures from various molecules together, generating a steric hindrance to further folding.

\begin{figure}[ht]
\par\centering
\resizebox*{0.9\textwidth}{!}
{\includegraphics[width=0.95\textwidth]{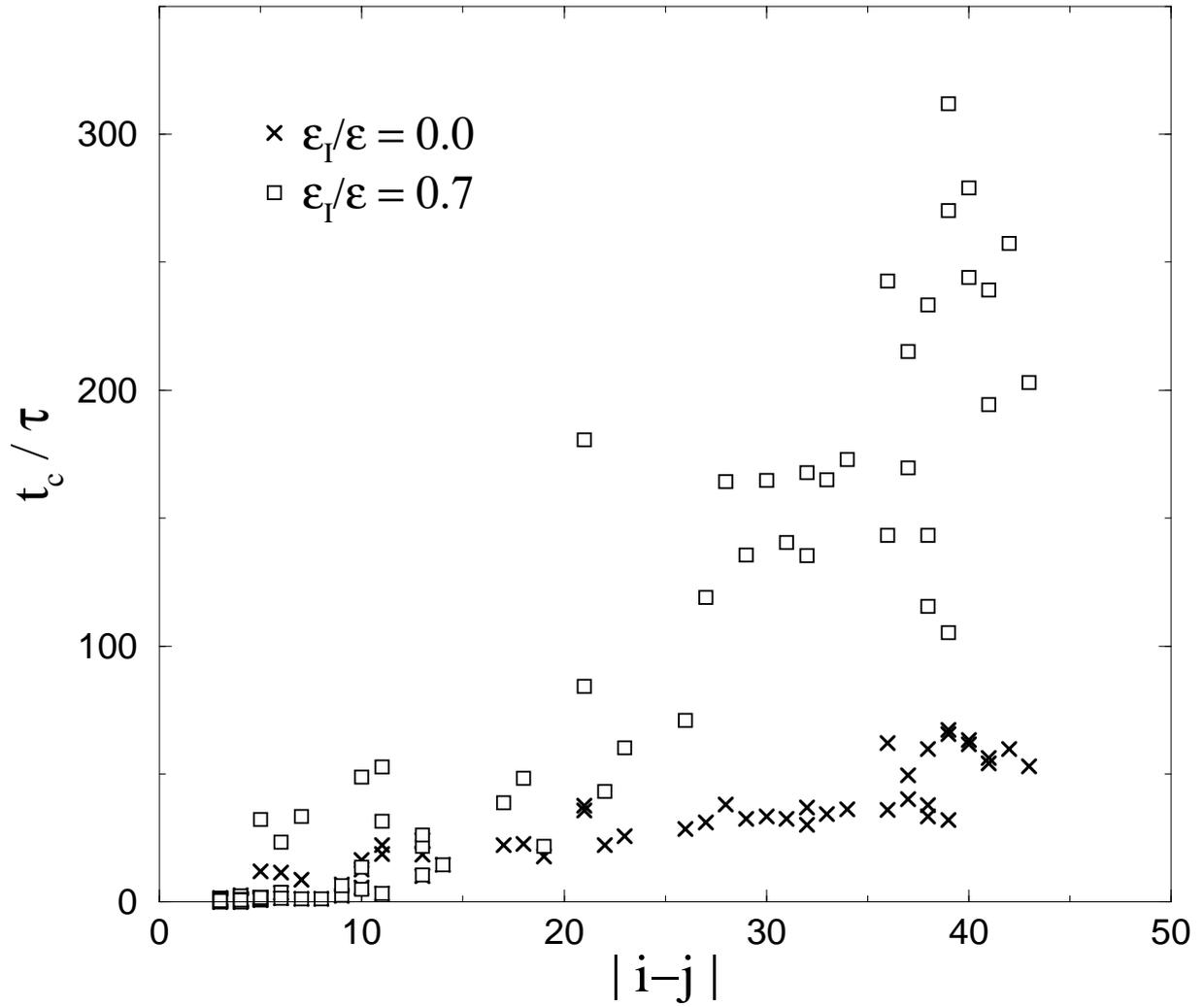}}
\par
\caption{The folding scenario for four proteins at $k_BT/\varepsilon$=0.3 for $\varepsilon _I/\varepsilon$=0 (crosses) and 0.7 (open squares).
}
\label{fig11}
\end{figure}

It should be noted that the presence of a substantial $\varepsilon _I$ also affects the folding scenario as seen in Figure \ref{fig11} for $\varepsilon _I/\varepsilon$=0 and 0.7 -- the case for which $t_{fold}$ varies by a factor of about 6.
It is seen that the inter-protein interactions delay establishment of the contacts as they compete with the internal interactions.
However, if the contact establishment times are normalized to $t_{fold}$ to bring out the relative changes in the structure of the scenario diagram then such relative times get shorter.

\section{Conclusions}

Our studies within the simple model employed here suggest that both confinement and crowding can affect the folding process of a protein.
The confinement related effects on  the folding time are weak unless there are attractive interactions between the proteins. 
It is also expected that increasing the number of proteins in a given sphere ($n$ bigger than 4) will eventually affect folding significantly. 
This point remains to be studied further.
The crowding effects may get enhanced further when one accounts for the hydrodynamic interactions in the system. 
One way to include these interactions is discussed in ref.\citep{Szymczak2007}.
It would be worthwhile to check other model proteins to check for any universalities in the behavior.

This paper is dedicated to Prof. Zbigniew Grzywna on his 60'th birthday and was motivated by a discussion with Nancy E. Levinger.
The work involved has been supported by the grant N N202 0852 33 from the Ministry of Science and Higher Education in Poland.

\newpage
% Fig 1

\newpage
% Fig 2

\newpage
% Fig 3

\newpage
% Fig 4

\newpage
%Fig 5

\newpage
% Fig 6

\newpage
% Fig 7

\newpage
% Fig 8

\newpage
%Fig 9

\newpage
% Fig 10

\newpage
% Fig 11

\end{document}